\documentclass[prl,superscriptaddress,showpacs,showkeys,twocolumn]{revtex4-1}
\usepackage{amsfonts,amsmath,amssymb,graphicx,epstopdf,verbatim}
\usepackage[english]{babel}
\newcommand{\kk}{\mathbf{k}}
\newcommand{\xx}{\mathbf{x}}

\begin{document}

\title{Dymamical Casimir emission from polariton condensates}
\author{Selma Koghee}
\affiliation{TQC, Universiteit Antwerpen, Universiteitsplein 1, B-2610 Antwerpen, Belgium}
\author{Michiel Wouters}
\affiliation{TQC, Universiteit Antwerpen, Universiteitsplein 1, B-2610 Antwerpen, Belgium}
\date{\today}
\begin{abstract}
We study theoretically the dynamical Casimir effect in an exciton-polariton condensate that is suddenly created by an ultrashort laser pulse at normal incidence. As a consequence of the abrupt change of the quantum vacuum, Bogoliubov excitations are generated. The subsequent evolution, governed by polariton interactions and losses, is studied within a linearized truncted Wigner approximation. We focus in particular on the momentum distribution and spatial coherence. The limiting behavior at large and small momenta is determined analytically. A simple scaling relation for the final condensate depletion as a function of the system parameters is found and the correlation length  is shown to depend linearly on the condensate depletion.
\end{abstract}

\maketitle

Interacting systems possess, as a profound consequence of quantum mechanics, a rich ground state structure. An immediate implication is the dynamical Casimir effect: when some parameter is abruptly varied, the system no longer finds itself in the new ground state, causing the appearance of excitations. Since the original idea of generating photons by accelerating two mirrors with respect to each other \cite{moore70,fulling76}, the dynamical Casimir effect has been studied in contexts as diverse as particle production in a time dependent cosmological background  \cite{birrell84,fedichev04,jain07}, superconducting circuits \cite{wilson11,lahteenmaki13} and ultracold atomic gases \cite{garay00,balbinot08,lahav10}. In the latter system, the nontrivial vacuum is that of the (weakly) interacting Bose gas, which features quantum depletion. 

A different realization of the weakly interacting Bose gas is a microcavity containing exciton polaritons \cite{iac_rmp}. Their advantage over atomic gases is that exciton polaritons have a photonic component, offering great flexibility for optically creating them in a specific state. This feature was e.g. exploited in the experiments on polariton superfluidity, where the polariton velocity could be varied by simply changing the angle of incidence of the excitation laser \cite{polsf}.

This advantage has recently led to several proposals to utilise exciton polaritons for the experimental observation of Hawking radiation \cite{hawking_mal,hawkingiac}, a phenomenon intimately related to the dynamical Casimir effect. In this Letter, we analyze a simpler manifestation, directly related to the quantum depletion of the interacting Bose gas. We propose to create polaritons by a pulsed laser at normal incidence, which results in a state with all particles at zero momentum. Since this is not the true ground state of the many body system, quasi-particles are expected to appear. In other terms, by the sudden creation of a coherent polariton population, the vacuum is changed from the trivial one to the Bogoliubov vacuum, leading to dynamical Casimir emission. The change of the dispersion due to the presence of a condensate was probed in the experiments of Kohnle {\em et al.} \cite{kohnle12} by a four wave mixing pump probe experiment. We suggest here to study the dynamics without the probe pulse.

An important aspect of polariton physics in actual experiments is the finite polariton life time. This brings the system in an inherently out of equilibrium situation and limits the possibilities to observe quantum effects. On the other hand, it offers a unique laboratory to study the interplay and competition between quantum correlations and dissipation, that may teach us lessons that are relevant for cosmological studies \cite{adamek13}. A basic question is how the achievable quantum depletion depends on the system parameters. We will show that within the Bogoliubov approximation, the dynamical quantum depletion of a one dimensional polariton wire \cite{wertz} scales linearly with a "blockade" parameter $U/\gamma$ \cite{verger}, where $\gamma$ is the line width and $U$ is the interaction energy of two polaritons that are localized within the condensate healing length.

The coupling between the polariton fluid and the outside world leads both to particle losses and to the entrance of quantum fluctuations into the system. This coupling can be described accurately with the master equation. For a quadratic Hamiltonian, an equivalent formulation in terms of the Wigner distribution of the polariton field can be constructed. This approach comes down to mapping the quantum master equation to a set of classical stochastic differential equations to sample the Wigner distribution function. The stochastic differential equation for the polariton field reads in the so-called truncated Wigner approximation \cite{cfield_review}
\begin{align}
i\hbar \, d \phi(\xx,t) =& \left[\epsilon \left(-i \vec{\nabla} \right) - i \frac{\gamma}{2} + \frac{g}{\Delta V} |\phi(\xx,t)|^2 \right] \phi(\xx,t) \, dt , \nonumber \\
&+ \sqrt{\frac{\gamma}{4}} dW(\xx,t),
\label{eq:wigev}
\end{align}
where $\epsilon(-i \vec{\nabla})$ gives the kinetic energy of a free particle, $g$ is the strength of the contact interaction, and $\Delta V$ is the volume of the grid cell necessary in the derivation of the truncated Wigner approximation. The stochastic term $dW$ is white Gaussian noise with amplitude variance $\langle dW^*(\xx,t) dW(\xx',t) \rangle = 2 dt \, \delta_{\xx,\xx'}$ and a random phase. Expectation values with respect to the Wigner distribution of the stochastic fields correspond to the symmetric operator averages, e.g. $\langle \hat\psi^\dag(\xx,t) \hat \psi(\xx',t') +  \hat \psi(\xx',t')  \hat\psi^\dag(\xx,t)\rangle = \langle \phi^*(\xx,t) \phi(\xx',t) \rangle_W$. This allows for a direct computation of all equal time correlation functions \cite{qnoise}.

The action of a resonant excitation pulse at $\kk=0$ can be included in Eq.\eqref{eq:wigev} by adding a pumping term $F_L(\xx,t)$ on the right hand side. However, if the pulse is very short, it can be replaced by a shift of the initial condition of the stochastic variable 
\begin{equation}
\phi(\xx,t=0)=\eta(\xx)+\int F_L(t') e^{i \epsilon(0) t'} dt' ,
\label{eq:incond}
\end{equation}
where $\eta(\xx)$ is the normally distributed uncorrelated vacuum noise with variance $\langle \eta^*(\xx)\eta(\xx') \rangle_W = 1/2 \, \delta_{\xx,\xx'}$. The subsequent dynamics is no longer influenced by the exciting laser, which merely fixes the initial condition. This situation is very different from both continuous wave resonant and nonresonant excitation schemes, where an interplay between driving and dissipation takes place at all times. Compared to the continuous wave excitation regime, the dynamics of the polariton gas in the present scheme resembles the standard Bose gas more closely.

\begin{figure}
	\includegraphics[scale=0.18, angle=0]{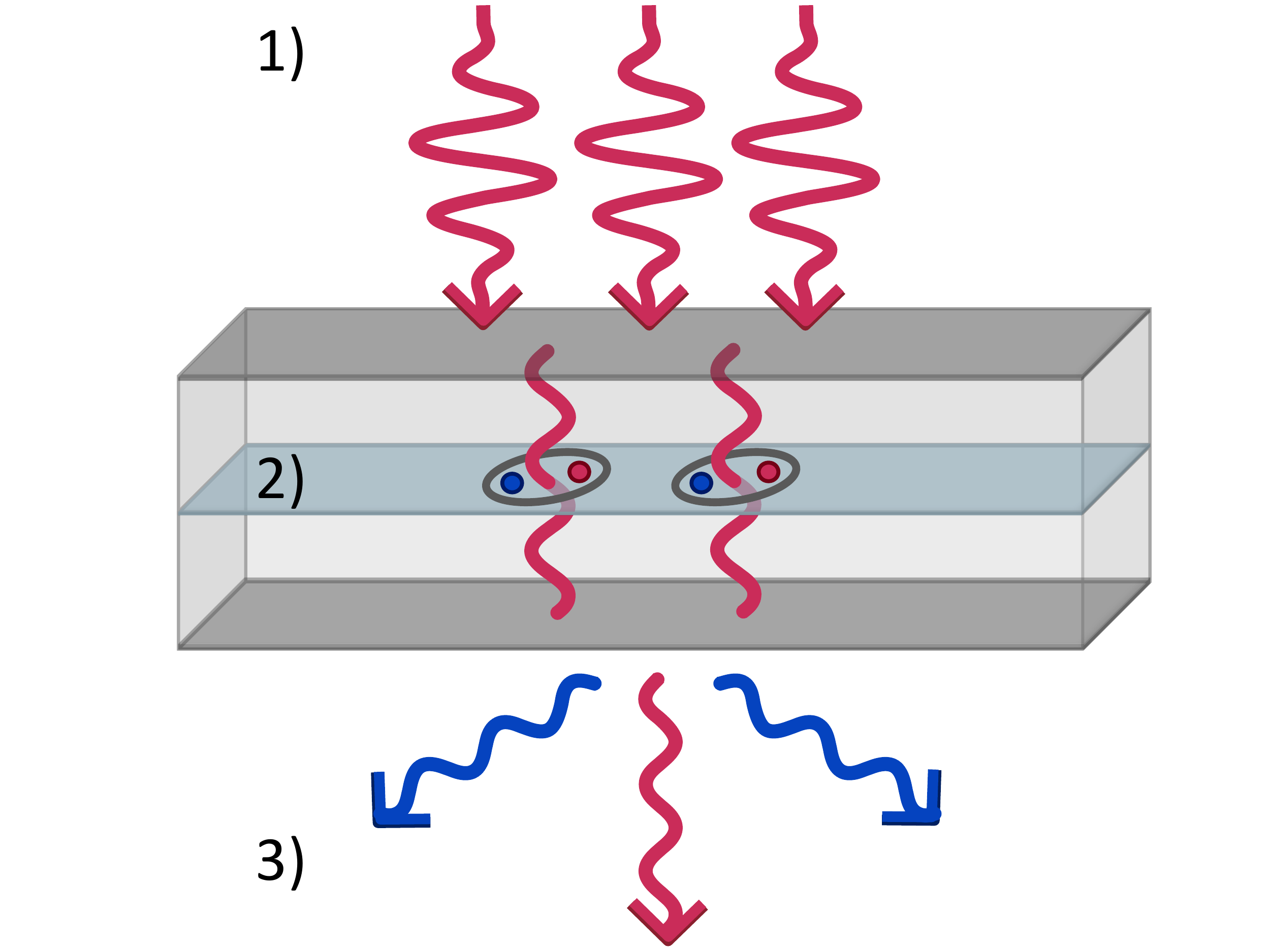}
	\caption{When a short laser pulse (1) resonantly creates a polariton condensate (2), a sudden change of trivial vacuum to the Bogoliubov one, results in the generation of dynamical Casimir emission (3) of Bogoliubov excitations.}
\end{figure}

Only in the case of a quadratic Hamiltonian, the mapping between the quantum and classical problem becomes exact. Since we will restrict our calculations to the Bogoliubov approximation, the Wigner forumulation does not introduce any further errors. Let us therefore decompose the polariton field in a condensate ($\kk=0$) and fluctuation component; $\phi(\xx,t)=\phi_c(t) + \delta \phi(\xx,t)$. In the equation of motion for the condensate, the effect of the interactions with the noncondensed particles and the effect of fluctuations are negligible within the Bogoliubov approximation. Thus, the time evolution of the condensate component is given by \cite{kohnle12}
\begin{equation}
\phi_c(t) = \sqrt{n_c(t)} \exp \left[ -\frac{\gamma t}{2 \hbar} - i \theta_c(t) \right],
\label{eq:phict}
\end{equation}
where the condensate density $n_c(t)=n^{(0)}e^{-\gamma t}$ decays exponentially and its phase is given by the time integrated blue shift $\theta_c(t)=g \int_0^t n(t') dt'$.
Up to first order, the motion of the fluctuations is governed by the familiar Bogoliubov matrix
\begin{equation}
B_\kk = \begin{pmatrix}
\epsilon(k) + gn_c(t)\, -i \gamma /2 & g n_c(t) \\
-g n_c(t)  & -\epsilon(k) - gn_c(t)\, -i \gamma /2
\end{pmatrix}
\label{eq:bogmat}
\end{equation}
as
\begin{equation}
i \hbar \, d \Phi_\kk = B_\kk(t) \Phi_\kk dt + \frac{\sqrt{\gamma}}{2} d\Xi_\kk(t),
\label{eq:bogdyn}
\end{equation}
where $ \Phi_\kk=[e^{-i \theta_c} \delta \phi(\kk)\;\; e^{i \theta_c} \delta \phi^*({-\kk})]^T$ and $d\Xi_\kk(t)=[dW(\kk,t) \; - dW^*(-\kk,t)]^T$, with the correlation of the noise in momentum space $\langle dW^*(\kk,t) \, dW(\kk',t)\rangle=2dt\, \delta_{\kk,\kk'}$.

Due to the time dependence of the Bogoliubov matrix, an analytic solution for the fluctuations 
$ \Phi_\kk$ is not straightforward to compute, but for small and large wave vectors, limiting expressions can be found analytically. Let us start with the limit $\kk \rightarrow 0$. In this case, the time dependence of the real part of the Bogoliubov matrix comes down to a rescaling with a factor $\exp(-\gamma t/\hbar)$, whereas the imaginary part is proportional to the identity matrix. Hence, the Bogoliubov matrices at different times commute with each other. The differential equations are then easily solved, yielding for the momentum distribution
\begin{align}
\lim_{k\rightarrow 0} n(\kk,t) &= \lim_{k\rightarrow 0} \langle \delta\phi^*(\kk) \delta\phi(\kk) \rangle_W- \frac{1}{2} \nonumber \\
 &= 2 \left(
\frac{g n^{(0)}}{\gamma}
\right)^2 e^{-2\gamma t/\hbar} \left(e^{\gamma t/\hbar} -\frac{\gamma t}{\hbar} -1 \right),
\label{eq:NK0}
\end{align}
where we have used the relation between the momentum distribution and the expectation value of the stochastic fields that sample the Wigner distribution $W$.
At large times, Eq. \eqref{eq:NK0} reduces to the simple scaling $n({\kk \rightarrow 0},t \rightarrow \infty) = 2 (g n^{(0)}/\gamma)^2 \exp(-\gamma t/\hbar)$.

For the other limit of large $k$, we can use the sudden approximation from \cite{carusotto10} to compute the number of Bogoliubov excitations that are created. At zero temperature, this approximation yields $n_B(k) = | v(k) |^2$, where $v(k)=\sqrt{[\epsilon(k)+g n^{(0)}]/[2 \hbar \omega_B(k)] -1/2}$ is the usual Bogoliubov $v$ coefficient which depends on the dispersion as $\hbar \omega_B(k)=\sqrt{\epsilon(k) \,[ \epsilon(k)+2g n^{(0)} ]}$. 
The polariton momentum distribution is related to the quasi-particle occupation as
\begin{align}
\langle \psi^\dag(\kk) \psi(\kk) \rangle = u(k)^2 \langle b^\dag(\kk) \, b(\kk) \rangle + v(k)^2 \langle b(-\kk) \, b^\dag(-\kk) \rangle \nonumber\\
+ u(k) v(k) \langle b(\kk) \, b(-\kk) \rangle + u(k) v(k) \langle b^\dag(-\kk) \, b^\dag(\kk) \rangle,
\label{eq:sudad}
\end{align}
The terms on the second line represent an oscillating contribution, that we will not take into account in the following. Using the initial occupation of the Bogoliubov excitations, and discarding losses one obtains for the momentum distribution $n(\kk,t)=2|u(k)|^2|v(k)|^2$. In the crudest approximation, losses can be included by an exponential decay:
\begin{equation}
n(k ,t) = \frac{1}{2} \left[ \frac{gn^{(0)}}{\hbar\omega_B(k)} \right]^2 e^{-\gamma t/\hbar}.
\label{eq:nklarg}
\end{equation}
The wave number $k_*$ where the cross-over between the small (\ref{eq:NK0}) and large (\ref{eq:nklarg}) momentum behavior takes place can be found by equating both expressions. In the more interesting limit where $g n^{(0)} \gg \gamma$, this yields
\begin{equation}
k_*(t)=\frac{\gamma}{2 \hbar} \sqrt{\frac{m}{g n^{(0)}}} \left[1- e^{-\gamma t/\hbar} \left(\frac{\gamma t}{\hbar} +1 \right) \right]^{-1/2}.
\label{eq:kstar}
\end{equation}
In the limit where the decay rate becomes much smaller than the interaction strength, this wave vector tends to zero and the expression (\ref{eq:nklarg}) becomes accurate down to very small wave vectors.

For the behavior of the momentum distribution around $k_*(t)$, we have not found an analytic formula, but the Bogoliubov dynamics (\ref{eq:bogdyn}) was solved numerically instead. To this purpose, we have introduced the Green's function of the time dependent linear system $G_\kk(t,t')$. The momentum distribution can then be expressed as
\begin{align}
&\langle \psi^\dagger(\kk,t) \psi(\kk,t) \rangle =  \\ 
&\int dt' 
\left\{|[G_\kk(t,t')]_{1,1}|^2 + |[G_\kk(t,t')]_{1,2}|^2 \right\}\frac{\delta(t')+1}{2} \; -\frac{1}{2} \;. \nonumber
\label{eq:momdis-gf}
\end{align}
The Green's function itself was computed by discretising time interval $[t=t_N,t'=t_1]$ in small steps $\Delta t$
\begin{equation}
G_\kk(t,t')=\Pi_{j=1}^N \exp[-i \Delta t B_\kk(t_j)].
\label{eq:greensfunction}
\end{equation}

\begin{figure}
	\includegraphics[scale=0.26, angle=0]{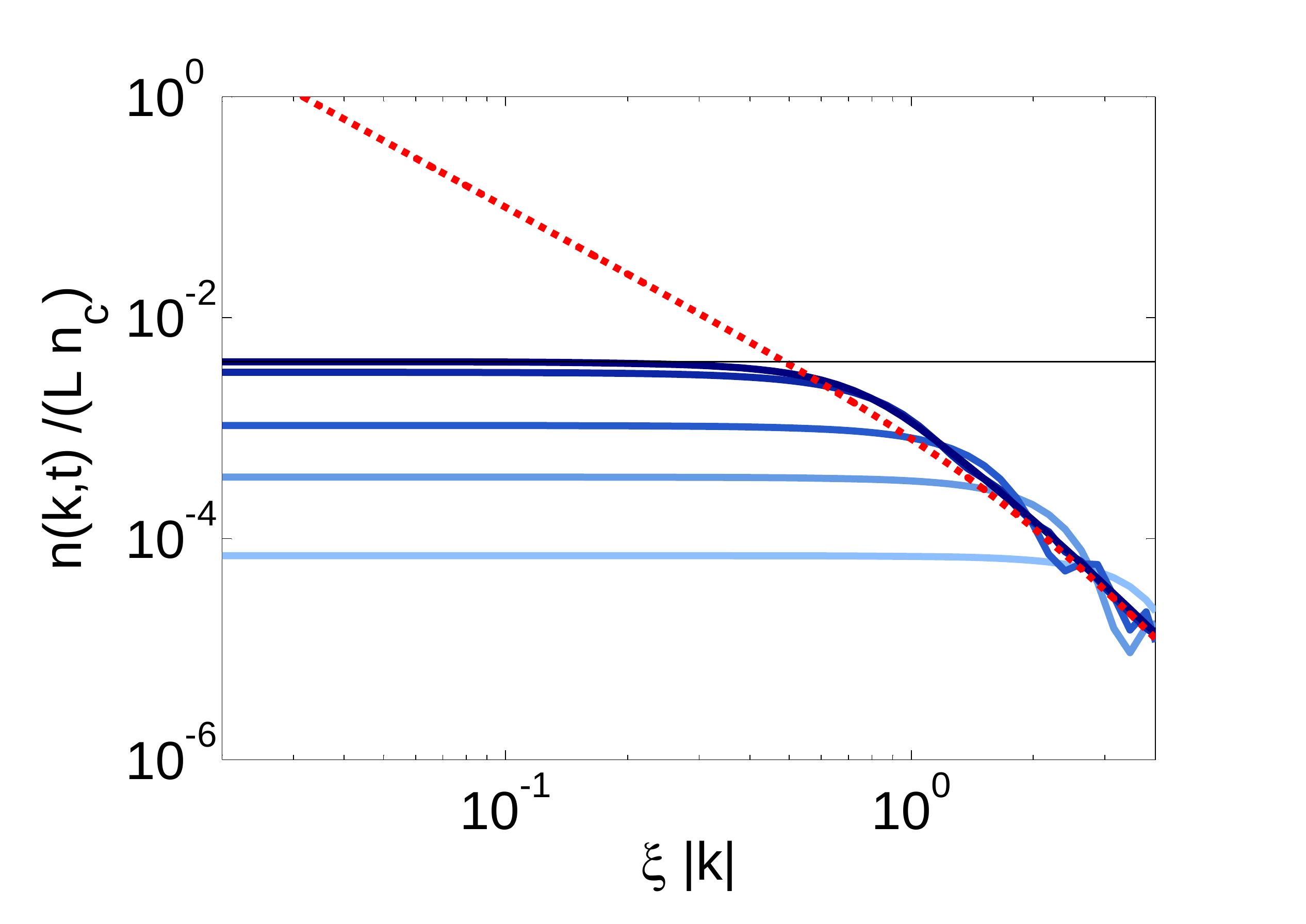} 
	\includegraphics[scale=0.26, angle=0]{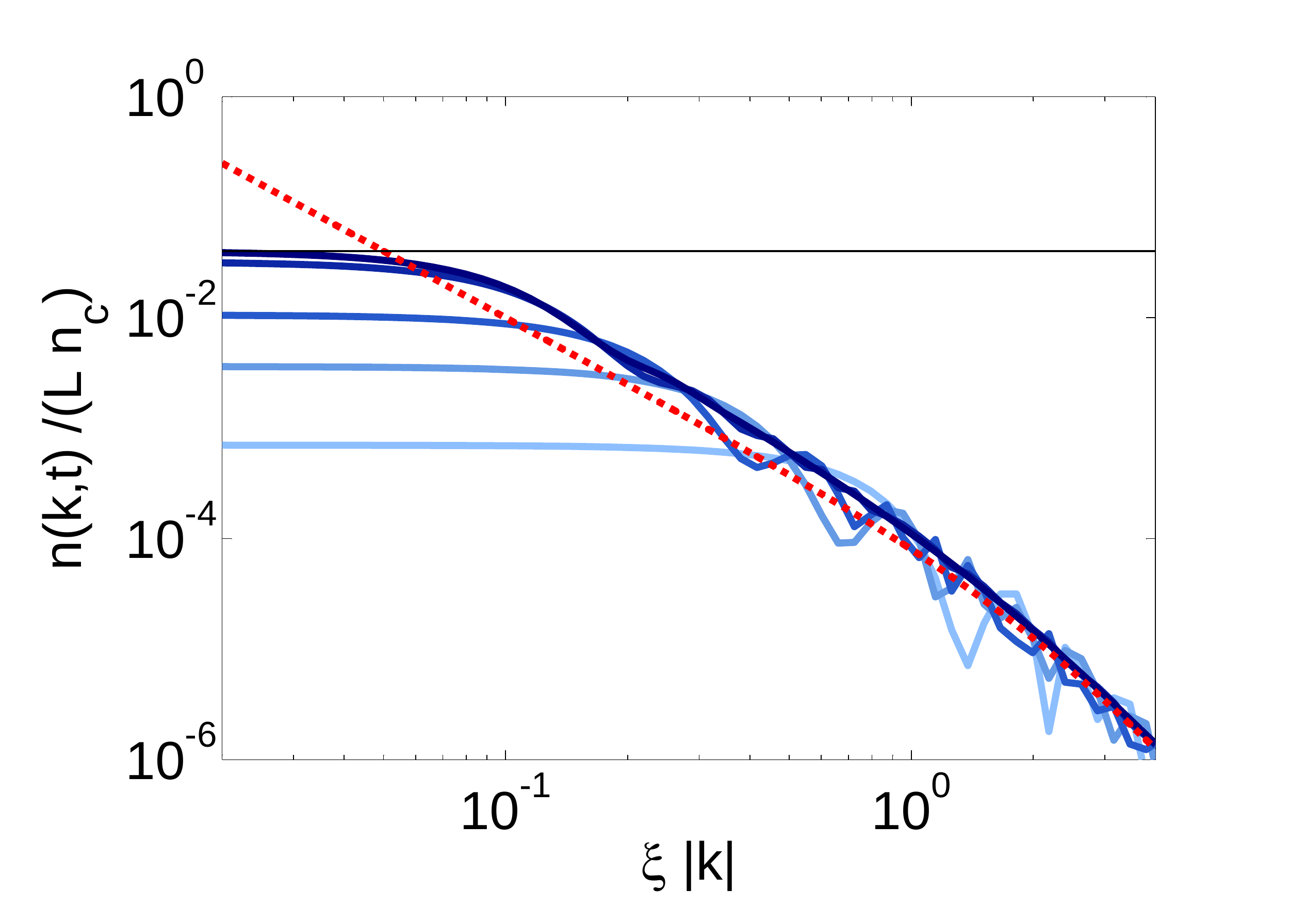}
	\caption{The momentum distribution $n(k)$ of the polariton gas due to the dynamical Casimir effect. The full lines show the numerical result for various times (t=0.2,0.5,1,3,8). The red dotted line represents the asymptotic result \eqref{eq:nklarg}; the thin black line is the prediction \eqref{eq:NK0} for small $\kk$ at large times. We have chosen $g=0.01 \, \mu {\rm m \, meV}, \; \gamma=0.05 \, {\rm  meV}, L= 100 \, \mu {\rm m}, \hbar=1, \; m=1$. Upper panel: $g n^{(0)}/\gamma=1$. Lower panel: $g n^{(0)}/\gamma=10$.}
	\label{fig:momdis}
\end{figure}

The momentum distributions obtained with this numerical scheme are presented in Fig. \ref{fig:momdis}. We have rescaled them to the condensate density in order to cancel the trivial exponential decay. The red dotted line represents the expression \eqref{eq:nklarg}, that is seen to match very well for sufficiently large momenta, justifying the sudden approximation in this regime. For the smaller interaction strength (a), the momentum distribution decays as $k^{-4}$, where for the larger interaction strength (b), its decay is dominated by the intermediate $k^{-2}$ region.  The black solid line represents the prediction \eqref{eq:NK0} for $t\rightarrow \infty$, which matches perfectly with the numerical result at late times. 

\begin{figure} 
\includegraphics[scale=0.25, angle=0]{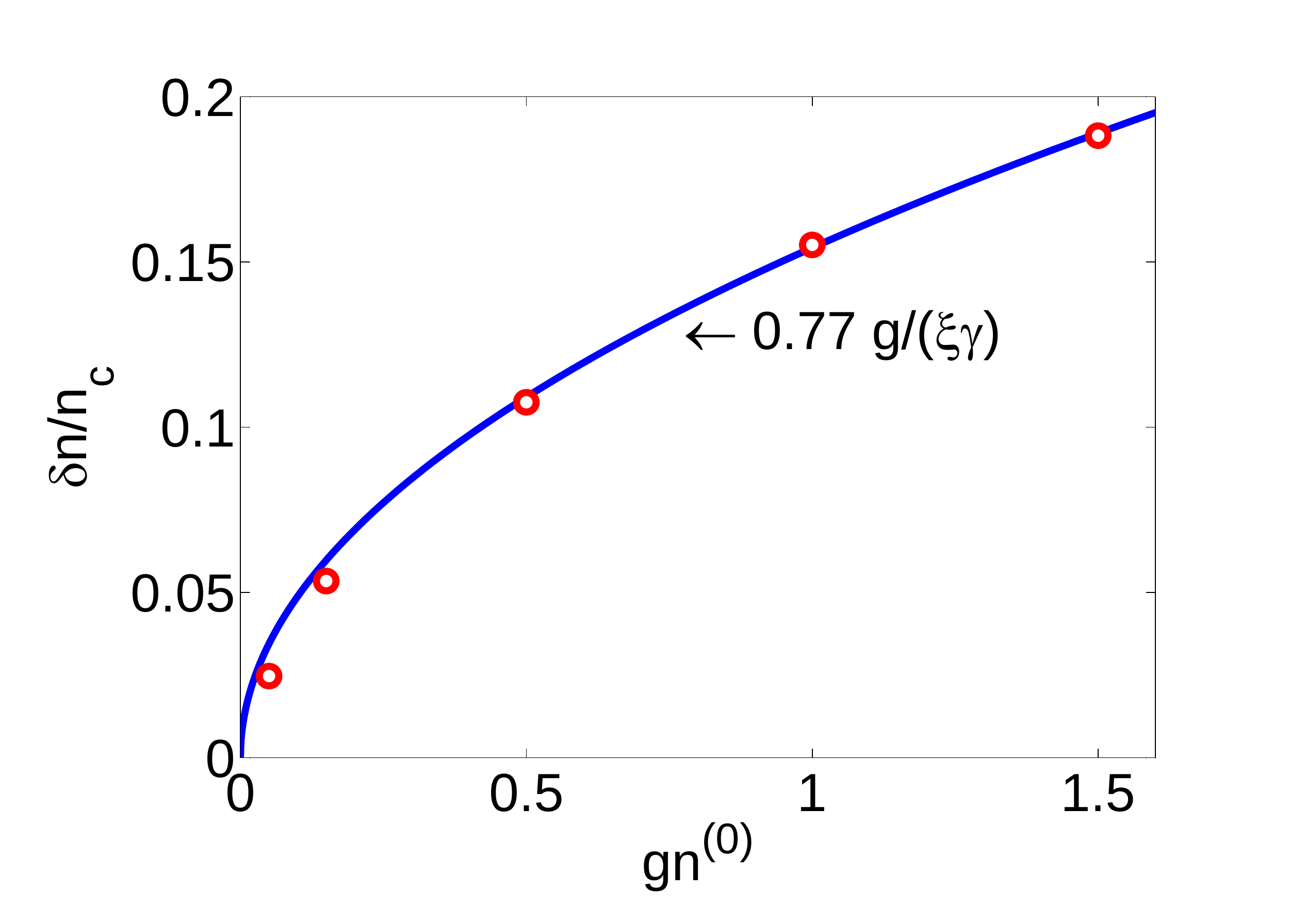}
\caption{The total quantum depletion as a function of the initial blueshift. The circles show the numerical result; the full line is the analytical prediction. We have taken an interaction strength $g=0.01 \mu {\rm m \, meV}$ and a decay rate $\gamma=0.05 \, \mathrm{meV}$.}
\label{fig:depletion}
\end{figure}

From the full momentum distribution, we can now compute the total number of particles that are produced by the dynamical Casimir effect. This result will depend on the dimensionality of the system. Here, we will evaluate the particle production for a one dimensional polariton wire, where quantum fluctuations are most important due to the high density of states at low momenta. At large times, the ratio of particles in nonzero momentum states to the condensate density $n_c(t)$ can be written as
\begin{equation}
\frac{\delta n(t)}{n_c(t)} = \frac{g^2 n^{(0)}}{\pi \gamma^2} \left[1- e^{-\gamma t/\hbar} \left(\frac{\gamma t}{\hbar} + 1 \right) \right] \int dk \frac{n(k,t)}{n(k \rightarrow 0,t)},
\label{eq:dn}
\end{equation}
where we have used \eqref{eq:NK0} for the number of particles created in the small momentum states and we have taken the continuum limit. 
By approximating the integrand in Eq. \eqref{eq:dn} by 1 for $k<k_*$ and by using the expression \eqref{eq:nklarg} for $k>k_*$, we find that the particle production scales in the limit $g n^{(0)} \gg \gamma$ as
\begin{equation}
\frac{\delta n(t)}{n_c(t)}
= C \; \frac{ g^2 n^{(0)}}{\gamma^2} \left[1- e^{-\gamma t/\hbar} \left(\frac{\gamma t}{\hbar} +1 \right) \right] k_*(t),
\label{eq:dnscale}\end{equation}
where the constant can be computed numerically from the full momentum distribution: $C\approx 1.5$. 
At large times, the depletion \eqref{eq:dnscale} can be written as $\delta n/n \approx 0.77 \;  g/(\xi \gamma)$, where $\xi=\hbar/\sqrt{mgn^{(0)}}$ is the healing length of the condensate immediately after its creation. This shows that the particle production is proportional to the ''blockade'' parameter $U/\gamma$ \cite{verger}, where $U=g/\xi$ is the interaction energy of two polaritons localized within the condensate healing length. Typical attainable values of the healing length are one micron, so that with an interaction strength $g=0.01 {\rm meV} \mu {\rm m}$, a linewidth of 0.05 meV is sufficient to obtain more than ten percent of quantum depletion. Figure \ref{fig:depletion} compares the analytical estimate \eqref{eq:dnscale} with the depletion computed from the full numerical momentum distribution. It turns out to be already accurate for moderate values of the parameter $gn^{(0)}/\gamma \geq 1$. Although large values of this parameter seem attainable with state of the art microcavities \cite{snoke}, this regime goes beyond the scope of the present work. When $gn^{(0)}/\gamma$ becomes too large, interactions between excitations can no longer be neglected, so the Bogoliubov approximation breaks down.

\begin{figure}
\includegraphics[scale=0.26, angle=0]{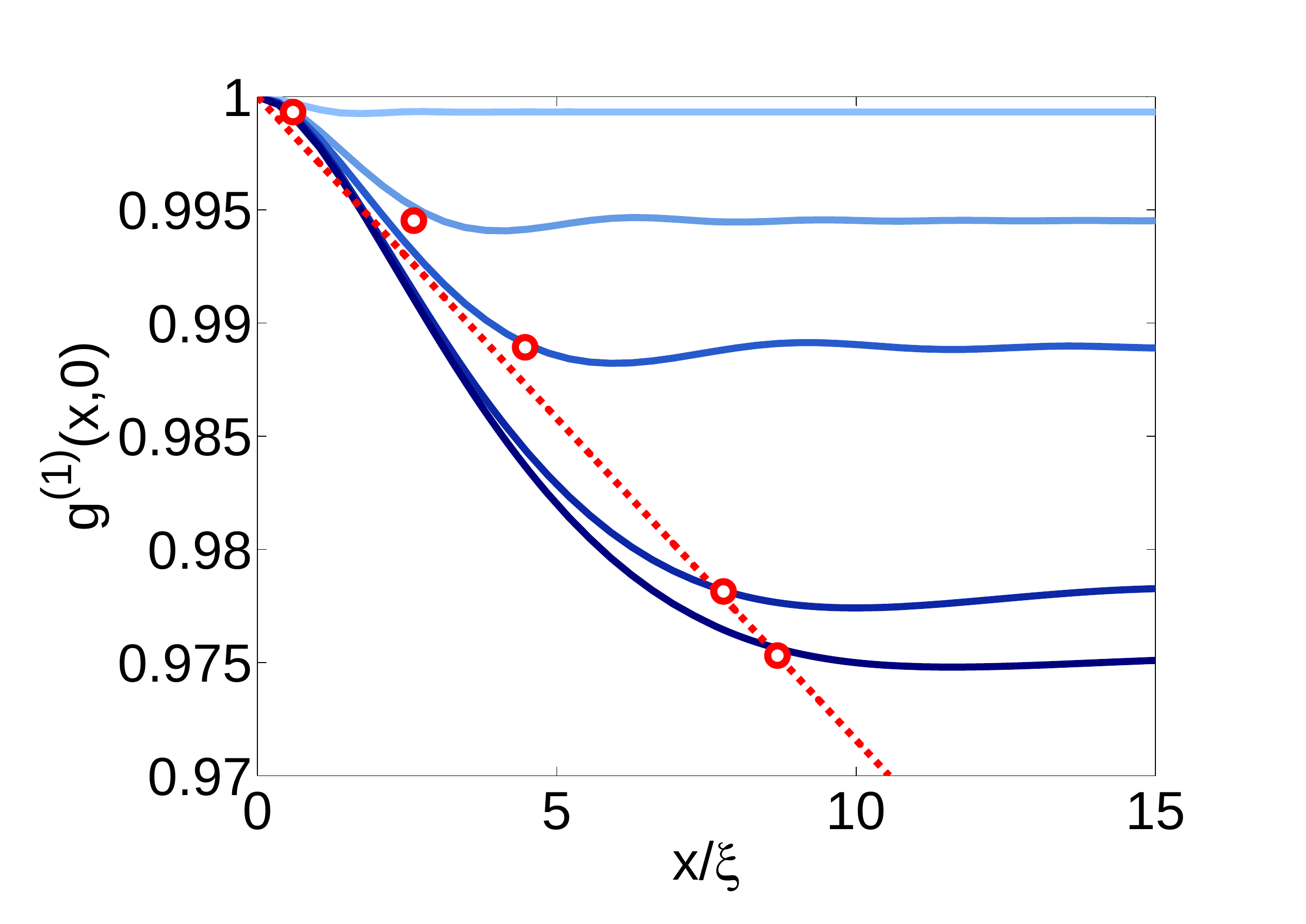}
\includegraphics[scale=0.26, angle=0]{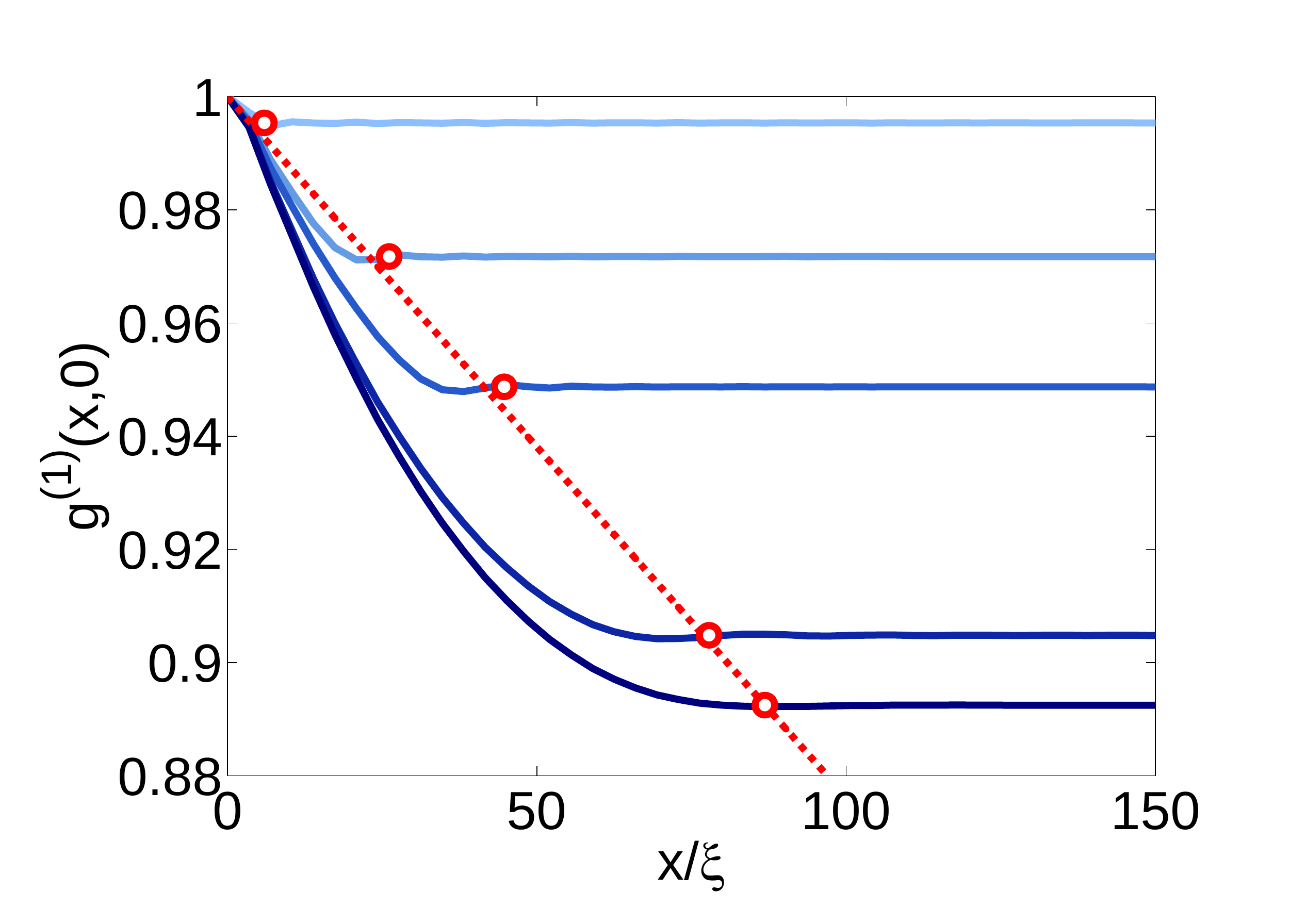}
	\caption{The spatial coherence $g^{(1)}(x)$ as a function of time ($t=0.1,0.5,1,3,8$), for the parameters as the corresponding panels in Fig. \ref{fig:momdis}. The red circles represent the analytical estimate of the correlation length $\ell_c(t)= 3.9/k_*(t)$. The red dotted line shows the linear relation between the correlation length and the depletion.}
	\label{fig:spatcoh}
\end{figure}

Experimentally, the quantum depletion is directly accessible through a measurement of the first order spatial coherence $g^{(1)}(x-x')=\langle \psi^\dag(x,t) \psi(x',t) \rangle/n_c(t)$. The fluctuations that are created because of parametric scattering lead to a drop in the first order coherence function equal to the quantum depletion: $g^{(1)}(x\rightarrow \infty)-g^{(1)}(x\rightarrow 0)=\delta n/n_c$.  We restrict again to a one dimensional wire, as we did for $\delta n/n_c$.

The numerical results for the first order coherence are presented in Fig. \ref{fig:spatcoh}. Since the gas is assumed to be created by a perfectly coherent laser, no fluctuations are present initially. 
The correlation length can be estimated from the relation $\ell_c(t)\propto 1/k_*(t)$. The red circles in Fig. \ref{fig:spatcoh} show this prediction to work well.  From this relation, it is seen that the coherence length scales at short times as $\ell_c(t) \propto 2 t \sqrt{gn^{(0)}/m} $. This is indeed what one expects for fluctuations that spread out at the speed of sound $c=\sqrt{g n_c(t)/m}$.
At long times, we obtain $\ell_c(t\rightarrow \infty) = 7.8 \, \sqrt{gn^{(0)}/m} \, \hbar/ \gamma$, where the constant $7.8$ was determined numerically. This means that the effective propagation time of the fluctuations is the life time $\tau=\gamma^{-1}$; their average speed is proportional to the initial sound velocity.  Finally, it is interesting to note in Fig. \ref{fig:spatcoh} the linear relation between the time dependent depletion and the coherence length: $\ell_c(t) \propto \delta n(t)/n_c(t)$. By combining Eqns. \eqref{eq:kstar} and \eqref{eq:dnscale}, this relation is readily verified analytically.

The dynamical Casimir generation of Bogoliubov excitations that we have studied here, is only the first process on the route of the Bose gas to thermal equilibrium. For long polariton life times, a second stage with collisions between Bogoliubov excitations should set in. This opens the prospect to explore the rich physics of the thermalization of quantum systems \cite{dynamicsrmp} in a new setting.

\end{document}